\documentclass{iopart}
\usepackage{graphicx,xcolor}
\usepackage[sort&compress,square,comma,numbers]{natbib}
\usepackage[utf8]{inputenc}
\usepackage{amsmath}
\usepackage{amssymb}
\usepackage{epic,eepic}
\usepackage{hyperref}

\DeclareMathOperator{\sgn}{sgn}

\renewcommand{\t}{\tilde}
\renewcommand{\o}{\bold}
\newcommand{\newL}{\;\;\; , \nonumber \\}
\renewcommand{\endL}{\;\;\; .}

\begin{document}
\title{Variational Cluster Approximation to the Thermodynamics of Quantum Spin Systems}
\author{S Filor, T Pruschke}
\address{Institute for Theoretical Physics, University of G\"ottingen, 37077 G\"ottingen, Germany\\}
\ead{filor@theorie.physik.uni-goettingen.de}

\begin{abstract}
We derive a variational cluster approximation for Heisenberg spin systems at finite temperature based on the ideas of  the self-energy functional theory by Potthoff for fermionic and bosonic systems with local interactions. Partitioning the real system into a set of clusters,
we find an analytical expression for  the auxiliary free energy, depending on a set of variational parameters defined on the cluster, whose stationary points provide approximate solutions from which the thermodynamics of spin models can be obtained.  We explicitly describe the  technical details of how to evaluate the free energy for finite clusters and remark on specific problems and possible limitations of the method. To test the approximation we apply it to the antiferromagnetic spin $1/2$ chain and compare the results for varying cluster sizes and choices of variational parameters with the exact Bethe ansatz solution.
\end{abstract}

\section{Introduction}
Magnetism is one of the fundamental phenomena in physical systems emerging from the interplay between Pauli's principle
and the Coulomb interaction. In condensed matter systems the macroscopic nature of the system together with the
crystal structure and details of the electronic bonding can lead to a huge variety of effects related to magnetism \cite{fazekas:99}, and a proper
theoretical description is still a major challenge. Loosely speaking one can distinguish two classes of magnetic materials,
viz itinerant and localized magnets. The latter can be found quite frequently in so-called Mott insulators based on transition
metal compounds \cite{imada:98} and can be well represented by models of localized spins interacting via an exchange interaction \cite{fazekas:99}. The simplest of such models is the Heisenberg model \cite{fazekas:99}
\begin{eqnarray}
	\o{H} \;=\; \sum_{i} h \, \bold{S}^{z}_{i} \,+\, \sum_{i j} \left [ J^{zz}_{i j} \, \bold{S}^{z}_{i} \bold{S}^{z}_{j} \,+\, \frac{1}{2} J^{-+}_{i j} \left( \bold{S}^{+}_{i} \bold{S}^{-}_{j} \,+\, \bold{S}^{-}_{i} \bold{S}^{+}_{j} \right) \right]  \;\; ,
	 \label{eq:1}
\end{eqnarray}
where we introduce a magnetic field and allow for an anisotropic exchange interaction when $J^{zz}_{i j} \neq J^{-+}_{i j}$. The physics of this model is well-known for the one-dimensional case \cite{takahashi:09}.
For dimensions $D\ge4$ one can apply Weiss mean-field theory which correctly describes the universal properties of phase transitions and can be used to at least qualitatively calculate physical quantities \cite{itzykson:89b}. 
For two and three dimensions, nearest-neighbor exchange
and simple lattices one can use highly efficient Monte-Carlo simulations to investigate the static and dynamic properties of the model \eqref{eq:1} \cite{sandvik:10}.

The situation is quite different when the magnetic interactions become frustrated, for example for the Heisenberg
model on triangular or Kagome lattices or 
in the presence of longer-ranged exchange interaction on square-lattices. In this case Quantum Monte-Carlo is plagued by a severe sign problem and reliable results for thermodynamic quantities cannot be obtained at low temperatures \cite{samson:93}. 
For certain situations, the sign problem can be circumvented by cleverly choosing the operator basis \cite{nakamura:98}, but in general it poses a severe restriction on Monte-Carlo simulations \cite{troyer:05}.
Alternative numerical approaches, like the density-matrix renormalization group \cite{schollwoeck:05} or variational Monte-Carlo 
\cite{shiba:93} are restricted to either one spatial dimension or ground-state properties only.

In the case of itinerant fermions Potthoff proposed a quite different ansatz,  the self-energy functional approach (SEFA)\cite{potthoff:03a}.  It is based
on the observation going back to Luttinger, Ward, Baym and Kadanoff \cite{baym:61}, that the free energy can be formally represented as functional of the fermionic Green function, with a non-trivial part called 
Luttinger-Ward-Baym-Kadanoff functional. The important aspect of this approach is that this latter functional
only depends on the structure of the interaction, but not on the kinetic energy of the fermions. 
This feature allows one to create well-defined approximations for models with strictly local interactions by
replacing the lattice by a collection of clusters, which can then be treated exactly. 
These variational cluster approaches (VCA) have been used to study various models for interacting fermion systems \cite{potthoff:03b,dahnken:04,aichhorn:04,potthoff:07, aichhorn:06, aichhorn:06a, arrigoni:09, senechal:08}. One can also derive the dynamical mean-field theory \cite{georges:review} 
within this framework \cite{potthoff:03a}. Koller and Dupuis later developed a  formulation of the VCA for systems
consisting of interacting bosonic particles, for example the Bose-Hubbard model \cite{koller:06}.

Potthoff's SEFA  rests on the representability of  the free energy as a unique functional of the single-particle Green function
respectively self-energy \cite{fetterwalecka}, with the contributions due to interactions being strictly separated from the 
non-interacting part. This property is in turn
based on a linked-cluster expansion for the free energy involving Wick's theorem, or equivalently relies on
standard Bose or Fermi commutation relations among the field operators constituting the non-interacting system \cite{fetterwalecka}. 
Spin operators, however, form another algebra, and so the standard formalism does not work. Nevertheless, due to the reasons mentioned above it would be interesting to have an analogous method for Heisenberg Hamiltonians since it would open a new possibility to tackle the problem of frustrated spin systems.  An additional category of models which cannot be treated easily within the standard Potthoff approach are those based on projected Hilbert spaces, like it is the case for the $t$-$J$ model relevant for high-$T_c$ superconductors  \cite{dagotto:94}. 
There do exist previous approaches where a diagrammatic perturbation theory for operators
with non-standard algebra was devised \cite{vaks:68,filor:10}, or a bosonization of the spin operators \cite{Ivanov:2004} introduced.  However, none of these approaches met their expectations.

Another class of systems which are not included in the standard VCA formulations are those with non-local
interactions as the necessary separation of local and non-local parts in the Hamiltonian is not possible
any more. Early attempts to include such interactions in theories like the dynamical mean-field theory
were usually based on scaling arguments \cite{si:96,smith:00} or assumptions about the structure how 
the fluctuation spectrum generated by these non-local interactions enters the free-energy functional \cite{georges:01}.
In 2005 Tong proposed a so-called extended variational cluster approximation (EVCA) for fermionic models with non-local interactions \cite{tong:05}. In this approach, a Luttinger-Ward-Baym-Kadanoff functional was explicitly constructed
from a fermionic coherent-state representation and tools of functional analysis were
used to establish a cluster approximation for such systems. Tong also suggested that such an approach could be used for a Heisenberg spin system \cite{tong:05}. 

In this paper, we will derive a formulation of the variational cluster approximation for spin systems 
and test them for a simple spin model. The paper is organized as follows. In the next section we will
introduce a coherent-state representation for spin operators. Within this formulation, we will derive an
expression which has the structure of a Luttinger-Ward-Baym-Kadanoff functional and which will serve
as basis to define an approximation based on a separation of the full system into clusters. This approach
will be tested for the spin $1/2$ Heisenberg chain in section \ref{sec:Results}. A
discussion of the features and deficiencies of the presented approach in section \ref{sec:Discussion} will
conclude the paper.

\section{Variational Cluster Approximation for Spin Models}\label{sec:SpinVCA}

\subsection{The Free Energy Functional} \label{sec:SpinVCA2.1}

The following discussion is based on the notation introduced by Tong \cite{tong:05}, which was developed for non-local electron-electron interactions. Since there are considerable differences for a Heisenberg Hamiltonian we have to treat the derivation of a spin variational cluster approximation (SVCA) thoroughly in this paper.  

As the starting point we use the spin path integral which can be derived by introducing spin-coherent states \cite{radcliffe:71,perelomov}. For a Hamiltonian of the form \eqref{eq:1} with isotropic interactions $J_{ij}$ and 
zero magnetic field the partition function can then be written in the following form \cite{wiegmann:88,fradkin:88}
\begin{eqnarray}
	Z &=& \int \, \prod_{i} \mathcal{D} \vec{s}_{i} \, e^{-S(\vec{s}_{i})} \;\;\; , \nonumber \\
	S(\vec{s}_{i}) &=& B(\vec{s}_{i}) 
	\,+\, \int_{0}^{\beta} \mbox{d} \tau \, \sum_{i j} J_{i j} \, \vec{s}_{i}(\tau) \vec{s}_{j}(\tau) \;\;\; .
	\label{eq:2}
\end{eqnarray}
The variables $\vec{s}_{i}$ are vectors of length $S$ which represent the quasi-classical path of the spins $\bold{S}$. The function $B(\vec{s}_{i})$ only depends on the structure of the manifold spanned by the possible paths 
and incorporates topological effects of a spin system. It represents the Berry phase \cite{radcliffe:71} and does not include any information on the interaction $J_{i j}$, which is solely present in the second term of the action \eqref{eq:2}. The term $B(\vec{s}_{i})$ is one
reason why the explicit evaluation of the spin path integral is rather complicated. However, to derive 
expressions for the free energy of a certain model and subsequently equations that allow to establish
a spin variational cluster approximation one only needs formal functional dependencies following from \eqref{eq:2}, i.e.\ the precise form of $B$ is not important. 

Let us define ${\o{J}}_{ij}(\tau - \tau') := J_{i j} \, \delta(\tau - \tau')$. As the exchange interaction
will later serve as variable for performing variations, we introduce an auxiliary field $\t{\o{J}}$, with  
the property $\t{\o{J}}={\o{J}}$ for the true physical system. The action is then formally written as a functional
\begin{equation}
	\t{S}[\vec{s}, \t{\o{J}}] \;=\; B(\vec{s}_{i})  
	\,+\, \int_{0}^{\beta} \mbox{d} \tau \int_{0}^{\beta} \, \mbox{d} \tau'  \, \sum_{i j} \, \vec{s}_{i}(\tau) \, \t{\o{J}}_{ij}(\tau - \tau')\, \vec{s}_{j}(\tau') \;\; .
	\label{eq:3}
\end{equation}
Henceforth, functionals will be denoted by a tilde which will be omitted if the corresponding quantity assumes its physical value. 

With this notation the partition function and  free energy become
\begin{eqnarray}
	\t{Z} [\t{\o{J}}] &=& \int \, \prod_{i} \mathcal{D} \vec{s}_{i} \, e^{-\t{S}[\vec{s}, \t{\o{J}}]} \;\; , \nonumber \\
	\t{F} [\t{\o{J}}] &=&  - \frac{1}{\beta} \ln \t{Z}[\t{\o{J}}] \;\; .
	\label{eq:4}
\end{eqnarray}
The form of the functional $\t{F}[\t{{\o{J}}}]$ depends on the structure of the exchange interaction $\t{{\o{J}}}$
only, but not its specific value.  With the help of the functional \eqref{eq:4}, the full spin-spin correlation function 
$\t{\o{\Pi}}$ is introduced as
\begin{eqnarray}
	\t{\o{\Pi}}_{i j}(\tau-\tau') &=& 
	\left< T_{\tau} \vec{\o S}_{i}(\tau) \vec{\o S}^{z}_{j}(\tau') \right>_{\t{S}} \; =\;- \beta \frac{\delta \t{F}[\t{\o{J}}]}{\delta \t{\o{J}}_{ji}(\tau-\tau')}\nonumber\\
	  &=& \frac{1}{\t{Z}} \, \int \, \prod_{i} \mathcal{D} \vec{s}_{i} \, \left( \vec{s}_{i}(\tau) \vec{s}_{j}(\tau') e^{-\t{S}[\vec{s}, \t{\o{J}}]} \right) \; .
	\label{eq:5}
\end{eqnarray}
Note that this definition is somewhat different from the standard one, which explicitly subtracts the expectation values of the spins and results in the connected correlation function. Although it is possible to formulate
the theory with this object, too, it turns out that the corresponding Hartree-like terms appearing in the action \eqref{eq:3}
lead to an additional set of constraints on local fields in the final formulation of the VCA for spin models, which are hard to satisfy for open spin systems. We therefore do not  follow this route further here.

In this paper we treat the more general case of a Hamiltonian \eqref{eq:1} where a finite magnetic field can be applied and which has the option of an anisotropic interaction. Here, a similar path integral representation can be derived using the spin-coherent states \cite{perelomov}. In contrast to the SU(2)-symmetric case we need to define two functional fields $\t{\o{J}}^{zz}$ and $\t{\o{J}}^{-+}$ to introduce a functional action $\t{S}[{s}^{\eta}, \t{\o{J}}^{zz}, \t{\o{J}}^{-+}]$ and the corresponding free energy functional. This is done in complete analogy to the above isotropic case. The integrands are now explicitly dependent on the spin vector components ${s}^{\eta}$. On the other hand it can be shown that the Berry phase $B(\vec{s}_{i})$ remains invariant \cite{perelomov, fradkin:88}. This is expected since this topological term does not depend on a specific Hamiltonian but rather on the paths of the single spins along the sphere.
 
The action $\t{S}[{s}^{\eta}, \t{\o{J}}^{zz}, \t{\o{J}}^{-+}]$ contains a local part 
\begin{equation}
	{S}^{loc}[{s}^{\eta}] = B(\vec{s}_{i}) \,+\,  \int_{0}^{\beta} \mbox{d} \tau \, \sum_{i} \, h {s}^{z}_{i}(\tau) \;\; ,
	\label{eq:5b}
\end{equation}
which consists of terms originating from the finite magnetic field and of the Berry phase. This is justified since the latter is composed of a sum over the individual spins. 

The longitudinal and transversal spin-spin correlation functions can now be defined as
\begin{eqnarray}
	\t{\o{\Pi}}^{zz}_{i j}(\tau-\tau') &=& - \beta \frac{\delta \t{F}[\t{\o{J}}^{zz}, \t{\o{J}}^{-+},\t{\o{J}}^{+-}]}{\delta \t{\o{J}}^{z z}_{ji}(\tau-\tau')} \nonumber \\
	&=& \frac{1}{\t{Z}} \, \int \, \prod_{i} \mathcal{D} {s}^{\eta}_{i} \, \left( {s}^{z}_{i}(\tau) {s}^{z}_{j}(\tau') e^{-\t{S}} \right) \nonumber \\
	&=&  \left< T_{\tau} {S}^{z}_{i}(\tau) {S}^{z}_{j}(\tau') \right>_{\t{S}} 
	\label{eq:6} \;\;\; , \\
	\t{\o{\Pi}}^{-+}_{i j}(\tau-\tau') &=&
	- \beta \frac{\delta \t{F}[\t{\o{J}}^{zz}, \t{\o{J}}^{-+},\t{\o{J}}^{+-}]}{\delta \t{\o{J}}^{-+}_{ji}(\tau-\tau')} 
	\nonumber \\&=& \left< T_{\tau} {S}^{-}_{i}(\tau) {S}^{+}_{j}(\tau') \right>_{\t{S}} \;\;\; . 
	\label{eq:7} 
\end{eqnarray}

There exists another correlation function, $\t{\o{\Pi}}^{+-}$, which  is given by an expression similar to \eqref{eq:7}. 
In the final expressions we will encounter traces over these two quantities, which then lead to identical contributions. For that reason we will not consider $\t{\o{\Pi}}^{+-}$ explicitly here. 

We will now use the functional relations  \eqref{eq:6} and  \eqref{eq:7} to derive the spin VCA equations. 
To keep the formulae simple, we use a compact notation without explicit reference to the components of
the functions, where appropriate. Note that in this case any trace also involves a sum over the different
longitudinal and transversal parts of the functions appearing.

\subsection{Luttinger-Ward Functional for Spin Systems}
We start by introducing a Legendre transformed auxiliary functional
\begin{eqnarray}
	\t{A}[\t{\o{\Pi}}] &=&  \t{F}[\t{\o{J}}] \,-\, \Tr \left(\frac{\delta \t{F}[\t{\o{J}}]}{\delta \t{\o{J}}} \, \t{\o{J}} \right) \nonumber \\ 
	&=& \t{F}[\t{\o{J}}] \,+\, \frac{1}{\beta} \Tr \left(\t{\o{\Pi}} \, \t{\o{J}} \right) \;\;\; , 
	\label{eq:9}
\end{eqnarray}
where $\Tr$ denotes the trace over spatial indices, imaginary time and spherical components.
The derivatives of the functional $\t{A}$ are
\begin{eqnarray}
	\frac{\delta \t{A}[\t{\o{\Pi}}]}{\delta \t{\o{\Pi}}^{z}} &=& \frac{1}{\beta} \, \t{\o{J}}^{z}[\t{\o{\Pi}}^{z}, \t{\o{\Pi}}^{t}] \newL
	\frac{\delta \t{A}[\t{\o{\Pi}}]}{\delta \t{\o{\Pi}}^{t}} &=& \frac{1}{\beta} \, \t{\o{J}}^{t}[\t{\o{\Pi}}^{z}, \t{\o{\Pi}}^{t}]  \;\;\; ,
	\label{eq:11}
\end{eqnarray}
where for simplicity we denote the longitudinal $zz$ and transversal $(-+,+-)$ correlation functions by a single $z$ and $t$, respectively.
With the help of Eq.\ \eqref{eq:9} we can write the free energy functional as Legendre transform of the functional
$\t{A}[\t{\o{\Pi}}]$, i.e.
\begin{equation}
	\t{F}[\t{\o{\Pi}}] \;=\; \t{A}[\t{\o{\Pi}}]  \,-\, \frac{1}{\beta} \Tr \left(\t{\o{\Pi}} \, \t{\o{J}} \right) 
	\endL
	\label{eq:12}
\end{equation}
So far not much can be said about the properties of the auxiliary functional. Of course the goal is to eventually derive some sort of Luttinger Ward functional. To this end we need to introduce the concept of a self energy for
the correlation functions $\t{\o{\Pi}}$. 

A similar quantity is used for example in the extended Dynamical Mean-Field Theory (EDMFT) for fermionic models with non-local interactions \cite{si:96, smith:99}. Here, a generalized self-energy $\o{\Gamma}$ can in principle be derived for a two-particle correlation function by using a cumulant expansion \cite{smith:00}. 
The resulting structure can be written in the general form \cite{tong:05}
\begin{equation}
	\o{\Gamma} \,=\, \o{J} + \alpha \o{\Pi}^{-1} \;\;\; ,
	\label{eq:13}
\end{equation}
with $\o{J}$ being a matrix consisting of the interaction parameters of the model and $\alpha$ some constant introduced to control the analytical properties of the approach. A typical choice is $\alpha=1/2$, which is also used by Tong \cite{tong:05}. 
However, in the case of spin models such a derivation does not readily exist, but one can nevertheless \emph{define} a self-energy of the
form \eqref{eq:13} from analogy arguments \cite{smith:00,georges:01}. This definition for $\o{\Gamma}$ is sensible because it allows to introduce a Luttinger-Ward functional with respect to correlation functions which has the same structure as the standard functional for single-particle Green functions \cite{tong:05, potthoff:2006}. 

On this level, the nature of the quantity $\o{\Gamma}$ seems somewhat artificial. 
Interestingly we can give it a well-defined meaning using the spin diagram technique introduced by Vaks, Larkin and Pikin \cite{vaks:68} and further developed by Izyumov and Skryabin \cite{izyumov:88}. 
It is a perturbative approach with respect to the spin exchange interaction, and
one can formally resum the diagrams to find relations similar to Dyson's equation, which in the present
context are called Larkin's equations \cite{izyumov:88}. They can be written in matrix notation as \cite{izyumov:88, pikalev:69} 
\begin{equation}
	\o{\Pi}^{\xi} \;=\; \o{\Sigma}^{\xi} \,+\, \o{\Sigma}^{\xi} \, \o{J}^{\xi} \, \o{\Pi}^{\xi} \label{eq:14} \;\; ,
\end{equation}
where $\xi$ stands for $z$ or $t$. The entries of the self-energy matrix $\o{\Sigma}$ represent the collection of all diagrams in the expansion of the correlation function that are irreducible with respect to one interaction line. This is a conceptual difference to the diagrammatic definition of the usual self-energy and responsible for
the slightly different structure of Eq.\ \eqref{eq:14}. It also means that Larkin's equation must not be
identified with a Dyson equation of standard perturbation theory. Nevertheless we can 
formally rewrite Eq.\ \eqref{eq:14} as
\begin{equation}
	\left( \o{\Sigma}^{\xi} \right)^{-1} \,=\, \o{J}^{\xi} \,+\, \left( \o{\Pi}^{\xi} \right)^{-1} \;\; . \label{eq:15} 
\end{equation}
Comparing this result to the expression \eqref{eq:13}, we now see that the previously defined quantity 
$\o{\Gamma}$ corresponds to the inverse of Larkin's self-energy with $\alpha=1$. There is no need to provide an explicit expression for the Larkin self-energy in the present approach, so \eqref{eq:15} can be used as a suitable and reasonable spin self-energy.
From now on we will conveniently use $\o{\Gamma}^{\xi}$ for the inverse Larkin self-energy $(\o{\Sigma}^{\xi})^{-1}$. 

We can now proceed with the derivation of a Luttinger-Ward-like functional for spin systems.
Using Eqs.\ \eqref{eq:11} and \eqref{eq:12} one can show that the following functional derivatives
\begin{eqnarray}
	\frac{\delta}{\delta \t{\o{\Pi}}^{z}} \, \left( \beta \t{A}[\t{\o{\Pi}}] + \Tr \ln  \t{\o{\Pi}}^{z} \right)  &=&  \t{\o{J}}^{z}[\t{\o{\Pi}}^{z}, \t{\o{\Pi}}^{t}] + \left( \t{\o{\Pi}}^{z} \right)^{-1} \nonumber \\
	&=& \t{\o{\Gamma}^{z}}[\t{\o{\Pi}}^{z}, \t{\o{\Pi}}^{t}]   \newL
	\frac{\delta}{\delta \t{\o{\Pi}}^{t}} \, \left( \beta \t{A}[\t{\o{\Pi}}] + \Tr \ln  \t{\o{\Pi}}^{t} \right)  &=& \t{\o{J}}^{t}[\t{\o{\Pi}}^{z}, \t{\o{\Pi}}^{t}] + \left( \t{\o{\Pi}}^{t} \right)^{-1} \nonumber \\
	&=& \t{\o{\Gamma}}^{t}[\t{\o{\Pi}}^{z}, \t{\o{\Pi}}^{t}] \;\; ,
	\label{eq:16}
\end{eqnarray}
hold, where we used the self-energies \eqref{eq:15}. We now define a generalized Luttinger-Ward functional
according to
\begin{eqnarray}
	\t{\Phi}[\t{\o{\Pi}}] \;=\; \beta \t{A}[\t{\o{\Pi}}] \,+\, \Tr \ln  \t{\o{\Pi}} 
	\endL
	\label{eq:17}
\end{eqnarray}
Of course one is faced with the question about the nature of this functional. 
The standard Luttinger-Ward functional can be derived as the collection of all connected closed skeleton diagrams 
\cite{luttinger:60b}. A similar identification is not obvious for the formal definition \eqref{eq:17}, as the
quantities appearing there are not explicitly connected to a diagrammatic expansion. 
Yet, as Tong already pointed out for non-local fermionic interactions \cite{tong:05}, a $\t{\Phi}$ such as \eqref{eq:17} is closely related to the formal derivation of the Luttinger-Ward functional for the standard Hubbard model introduced by Potthoff \cite{potthoff:2006}. It can be shown that this Luttinger-Ward functional has several important properties, which we will discuss now with respect to \eqref{eq:17}.

First, the free energy of the system can be written as a functional of the correlation functions using $\t{\Phi}$. Equations \eqref{eq:12} and \eqref{eq:17} lead to
\begin{equation}
	\beta \t{F}[\t{\o{\Pi}}] = \t{\Phi}[\t{\o{\Pi}}] \,-\, \Tr \ln  \t{\o{\Pi}} \,-\,  \Tr \left(\t{\o{\Pi}}^{z} \; \t{\o{J}}^{z}[\t{\o{\Pi}}^{z}, \t{\o{\Pi}}^{t}]  
	\,+\, \t{\o{\Pi}}^{t} \; \t{\o{J}}^{t}[\t{\o{\Pi}}^{z}, \t{\o{\Pi}}^{t}] \right) \endL
	\label{eq:18}
\end{equation}
Secondly, the functional derivatives of the Luttinger-Ward functional with respect to the correlation function are given by the self-energy \eqref{eq:15}, i.e.
\begin{eqnarray}
	\frac{\delta \t{\Phi}[\t{\o{\Pi}}]}{\delta \t{\o{\Pi}}^{z}} &=& \t{\o{\Gamma}}^{z}[\t{\o{\Pi}}^{z}, \t{\o{\Pi}}^{t}] \newL
	\frac{\delta \t{\Phi}[\t{\o{\Pi}}]}{\delta \t{\o{\Pi}}^{t}} &=& \t{\o{\Gamma}}^{t}[\t{\o{\Pi}}^{z}, \t{\o{\Pi}}^{t}] \;\; ,
	\label{eq:19}
\end{eqnarray}
This property is readily shown using Eqs. \eqref{eq:16} and \eqref{eq:17}.
It defines the self-energies as functionals of the two correlation functions. When evaluated at the physical 
values $\o{\Pi}$, the functionals $\t{\o{\Gamma}}$ acquire their physical value in Larkin's sense.  

The third property the generalized Luttinger-Ward functional should have is that it is universal in the sense that it does not explicitly depend on the interaction parameters $J$. This is however a direct consequence of the definition of the free energy functional $\t{F}[\t{\o{J}}]$ in \eqref{eq:4} and \eqref{eq:12}. It is defined by the structure of the local action \eqref{eq:5b} and the form in which the $\t{\o{J}}$ are introduced, not their explicit values. The property is inherited by the functional derivatives of $\t{F}[\t{\o{J}}]$ and the Legendre transform $ \t{A}[\t{\o{\Pi}}]$. Therefore the functional \eqref{eq:17} by construction does not depend on the specific interaction parameters.

\subsection{The Spin VCA Equations} \label{sec:SpinVCA2.3}

Even though the functional $\t{F}$ from Eq.\ \eqref{eq:18} was derived in a formal way and is not directly
based on a perturbative approach it can be used as starting point for approximations like the VCA. It was cast into a form with a structure which closely resembles the Baym-Kadanoff functional \cite{baym:61}, and its constituent parts are well defined. The generalized 
Luttinger-Ward functional has the postulated properties, and a self-energy can be defined in a 
Larkin irreducible sense.

Following the idea of Potthoff's original approach and Tong's work we now rewrite the free energy as a 
functional of the spin self-energies $\o{\Gamma}$ \eqref{eq:15}. For this step we need to introduce a Legendre transform of the Luttinger-Ward functional \eqref{eq:17}
\begin{eqnarray}
	\t{P}[\t{\o{\Gamma}}] &=&  \t{\Phi}[\t{\o{\Pi}}]    
	\,-\, \Tr \left(\frac{\delta \t{\Phi}[\t{\o{\Pi}}^{z}, \t{\o{\Pi}}^{t}]}{\delta \t{\o{\Pi}}^{z}} \, \t{\o{\Pi}}^{z} \right) 
	\,-\, \Tr \left(\frac{\delta \t{\Phi}[\t{\o{\Pi}}^{z}, \t{\o{\Pi}}^{t}]}{\delta \t{\o{\Pi}}^{t}} \, \t{\o{\Pi}}^{t} \right) \nonumber \\
	&=&  \t{\Phi}[\t{\o{\Pi}}] \,-\, \Tr \left(\t{\o{\Gamma}}\, \t{\o{\Pi}}\right)
	\endL
	\label{eq:21}
\end{eqnarray}
The functional derivates with respect to the self-energies are
\begin{eqnarray}
	\frac{\delta \t{P}[\t{\o{\Gamma}}]}{\delta \t{\o{\Gamma}}^{z}} &=& - \, \t{\o{\Pi}}^{z}[\t{\o{\Gamma}}^{z}, \t{\o{\Gamma}}^{t}] \newL
	\frac{\delta \t{P}[\t{\o{\Gamma}}]}{\delta \t{\o{\Gamma}}^{t}} &=& - \, \t{\o{\Pi}}^{t}[\t{\o{\Gamma}}^{z}, \t{\o{\Gamma}}^{t}]  \endL
	\label{eq:22}
\end{eqnarray}
These equations can be seen as defining the $\t{\o{\Pi}}$ as functionals of  $\t{\o{\Gamma}}$, i.e.\ we can write the free energy \eqref{eq:18} with the help of equation \eqref{eq:21} as a functional of these self-energies according to
\begin{eqnarray}
	\beta \t{F}[\t{\o{\Gamma}}] &=& \t{P}[\t{\o{\Gamma}}] \,-\, \Tr \ln \t{\o{\Pi}} 
	-\, \Tr \left(\t{\o{\Pi}}\, \t{\o{J}} \right) \,+\, \Tr \left(\t{\o{\Gamma}} \, \t{\o{\Pi}} \right)
	\endL
	\label{eq:23}
\end{eqnarray}
where all quantities are now to be taken as functionals of $\t{\o{\Gamma}}$. The last two terms can be absorbed into $\t{P}[\t{\o{\Gamma}}]$ by using the defining relations \eqref{eq:16}, and we end up with the expression
\begin{equation}
	\beta \t{F}[\t{\o{\Gamma}}] \;=\; \t{P}[\t{\o{\Gamma}}] \,+\, \Tr \ln \left(\t{\o{\Gamma}}^{z} \,-\, \t{\o{J}}^{z}\right) \,+\, \Tr \ln \left(\t{\o{\Gamma}}^{t} \,-\, \t{\o{J}}^{t}\right) \endL
	\label{eq:24}
\end{equation}

A central feature of the fermionic Luttinger-Ward functional not invoked yet is that it is stationary with respect to the corresponding single-particle Green function \cite{baym:61}. 
A similar stationarity condition holds for the functional (\ref{eq:24}) 
\begin{equation}
	\beta \t{F}_{SVCA}[\t{\o{\Gamma}}] \;=\; \t{P}[\t{\o{\Gamma}}] \,+\, \Tr \ln \left(\t{\o{\Gamma}}^{z} \,-\, \o{J}^{z}\right) \,+\, \Tr \ln \left(\t{\o{\Gamma}}^{t} \,-\, \o{J}^{t}\right) \;\; ,
	\label{eq:25}
\end{equation}
at the physical point $\t{\o J}^{\xi}=\o{J}^{\xi}$. We will refer to the functional \eqref{eq:25} as spin VCA (SVCA) functional
in the following.
For $\t{F}_{SVCA}$ we find
\begin{eqnarray}
	\left.\frac{\delta \t{F}_{SVCA}[\t{\o{\Gamma}}]}{\delta \t{\o{\Gamma}}}\right|_{\t{\o{\Gamma}}=\o{\Gamma}} &=& 0 \;\;\; , \nonumber \\
	 \t{F}_{SVCA}[\t{\o{\Gamma}}=\o{\Gamma}] &=& F \endL
	 \label{eq:26}
\end{eqnarray}

We previously argued that the Luttinger-Ward functional \eqref{eq:17} is universal in the sense that it does not 
depend explicitly on the interaction functionals $\t{\o{J}}$. This property is inherited by $\t{P}$ 
which means that its functional form is the same for \eqref{eq:24} and \eqref{eq:25}. 
We can thus eliminate it by subtracting the two free energy functionals, to arrive at
\begin{equation}
	\beta \t{F}_{SVCA}[\t{\o{\Gamma}}] = \beta \t{F}[\t{\o{\Gamma}}] 
	\,+\, \Tr \ln \left(\frac{\t{\o{\Gamma}}^{z} \,-\, \o{J}^{z}}{\t{\o{\Gamma}}^{z} \,-\, \t{\o{J}}^{z}}\right) 
	\,+\, \Tr \ln \left(\frac{\t{\o{\Gamma}}^{t} \,-\, \o{J}^{t}}{\t{\o{\Gamma}}^{t} \,-\, \t{\o{J}}^{t}}\right) \endL
	\label{eq:27}
\end{equation}
This functional $\t{F}_{SVCA}[\t{\o{\Gamma}}]$ is the equivalent of a Potthoff functional for a Heisenberg spin system. Although it was derived in a rather abstract way, it nevertheless can serve as starting point for a SVCA. 

To this end one has to choose a proper reference system, i.e.\ a system of exactly solvable clusters that shares the  
local Hamiltonian $H_{loc}$ with the original system, while differing in the interactions $\o{J}$. 
The functionals $\t{\o{J}}$ , $\t{\o{\Gamma}}$ and $ \t{F}$ in \eqref{eq:27} are then replaced by the corresponding quantities of 
this reference system, their particular values depending on the interactions $\o{J}_{c}$ of the cluster. With a
specific choice of these parameters we restrict the space of spin self-energies and obtain the expression 
\begin{equation}
	\beta F_{SVCA}(\o{J}_{c}) = \beta F_{c} \,+\, \Tr \ln \left(\frac{\o{\Gamma}_{c}^{z} 
	\,-\, \o{J}^{z}}{\o{\Gamma}_{c}^{z} \,-\, \o{J}_{c}^{z}}\right) 
	+\, \Tr \ln \left(\frac{\o{\Gamma}_{c}^{t} \,-\, \o{J}^{t}}{\o{\Gamma}_{c}^{t} \,-\, \o{J}_{c}^{t}}\right) \endL
	\label{eq:28}
\end{equation}

\begin{figure}[tb]
\vspace*{-3mm}
\begin{center}
\includegraphics[width=1\linewidth]{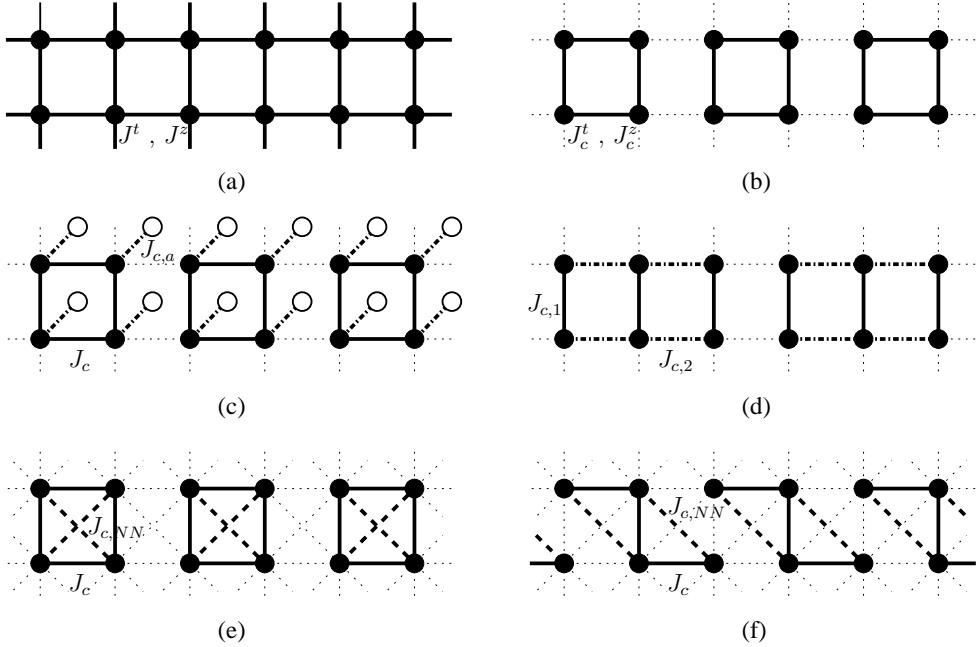}
  \end{center}
\caption{(a) The original square lattice with interactions $J^{z}$ and $J^{t}$. (b) An example of a reference system used in the SVCA consisting of simple four-spin clusters with the possible variational parameters $J_{c}^{z}$ and $J_{c}^{t}$ which can be varied independently. (c) The same reference system as (b) but with additional auxiliary sites interacting through the variational parameter $J_{c,a}$. (d) A six-spin cluster system which demonstrates the possibility of two different parameters $J_{c,1}$ and $J_{c,2}$ connecting different sites. (e), (f) Two examples of clusters usable in the case of an additional next-nearest-neighbor interactions in the original Hamiltonian. Here the two parameters $J_{c}$ and $J_{c,NN}$ can also be varied independently.}
  \label{fig:1}
\end{figure}

Examples of suitable cluster systems for a square lattice can be found in figure \ref{fig:1}, including possible variational parameters. 
Note that like in the fermionic approaches one has a rather large freedom regarding the clusters and parameters \cite{potthoff:2006b}.
The exchange interactions connecting different spins as well as the components $J_{c}^{z}$ and $J_{c}^{t}$ can in principle be varied independently from one another. Also, additional sites can be added to the boundary of the cluster, and so on. 
However, to keep the computation practicable one usually restricts the variational space to a reasonable set of parameters. 

Similar to the usual SEFA for fermionic systems, the task then consists in finding a stationary point of the SVCA equation with respect to the chosen $\o{J}_{c}$. Therefore Eq.\ \eqref{eq:28} needs to be evaluated explicitly. Details on the actual technical implementation of this step can be found in \ref{app:SVCA}. 

An extremal point determined in this way represents an approximation to the stationarity condition \eqref{eq:26}.
The results will of course depend on the specific choice of the cluster system and its parameters. In general one will expect that the approximation becomes better and less dependent on the actual selection with increasing cluster size.
Similar to the SEFA for fermions it is hard to define a limit where the method becomes exact. For our theory this is the special case $\o{J} \rightarrow 0$, i.e. for a model of decoupled spins. Here, the SVCA becomes exact  for a reference system of single-site clusters. 
\\

Before we test the approach on a specific model two problems have to be discussed. As pointed out in 
\ref{app:SVCA}, there exists the possibility to obtain complex frequency poles in the approximate physical correlation functions. Since this should not
be the case for a system in thermal equilibrium we conclude that the SVCA method can possibly break down and may not lead to meaningful results for a certain reference cluster with a specific set of parameters. 
Although such a breakdown need not be related to physical effects in any way, a possible interpretation could be that it signals a phase transition. The finite imaginary parts of the poles do not appear suddenly, but usually evolve continuously from the real axis to the imaginary axis by crossing the origin. 
However, as  discussed in \ref{app:SVCA}, one has to demand that for consistency reasons there is no pole at zero frequency \cite{koller:06}. In bosonic systems a non-vanishing excitation at the complex frequency origin signals the development of a condensed 
phase. 

For example, in VCA calculations for the Bose-Hubbard model, where breakdowns are also encountered, they are identified as a signal for the appearance of a superfluid phase \cite{koller:06}. This can be amended by formulating the VCA using a 'pseudo-particle' approach \cite{knap:11} or more rigorously within the Nambu formalism \cite{arrigoni:11}. Both methods give valuable insight how condensed phases of bosonic systems can be treated in a variational cluster approximation.
Although the physics of the spin operators is entirely different, it is tempting to invoke a similar
interpretation for the breakdown of the approximation behind the SVCA, viz that a different magnetic phase evolves for which a certain cluster structure is not suited any more. If this is true, one should in principle be able to set up a generalized cluster Hamiltonian that respects such a phase transition.

One further comment has to be added. The breakdown we just discussed takes place below certain temperatures, which means that often $T=0$ cannot be reached. This is the reason why it is advisable to evaluate the free energy \eqref{eq:28} using a method suitable for finite temperatures as described in \ref{app:SVCA}. 
\\

A second problem for the SVCA is that it has a limitation on possible variational parameters of the reference 
system. In the derivation of the SVCA free energy it was necessary to assume that the formal structure of the 
local action \eqref{eq:5b} respectively the corresponding Hamiltonian describing the reference system
is the same as for the model under consideration. The definition of the free energy functional \eqref{eq:12}, its Legendre transform and the generalized Luttinger-Ward functional critically depend on this property. In particular, we could not set up the SVCA equation \eqref{eq:27} by eliminating $\t{P}$ otherwise.

So the local action \eqref{eq:5b} needs to be unchanged which is naturally fulfilled for the Berry phase term since it is independent of the actual Hamiltonian. The only limitation here would be that the magnitude of the spins remains the same in the reference cluster system, which is natural and reasonable in any case. The second term of \eqref{eq:5b} on the other hand imposes the restriction that the local magnetic fields $h$ are fixed during the variation. Also, no additional local fields can be included in the reference system. Only entries in the interactions $\o{J}^{\xi}$ are eligible as variational parameters. On the other hand, variational local fields have proven to be a valuable tool in conventional VCA approaches, necessary to study certain states and phases of a system 
\cite{dahnken:04,aichhorn:04,senechal:08,koller:06}. 
That we cannot use magnetic fields in such a way is a limitation of our SVCA method and it is currently not clear how to lift this restriction. 
The only possible solution at present is the introduction of local anisotropies like $(\o{S}^{z}_{i})^{2}$ in the cluster Hamiltonian. However, such terms give no benefit for the spin $1/2$ chain used in the next chapter.

We would like to point out that the above limitation is also the reason why we do not use the 
connected longitudinal correlation function in the SVCA. In this case, the local part \eqref{eq:5b} of the action would also include terms proportional to $\left<\o{S}^{z}_{i}\right>$. The requirement that these terms remain unchanged for the original and cluster 
system can in general not be met for the open spin clusters treated so far. 

\section{Results} \label{sec:Results}

As a model to test our method  we use the antiferromagnetic spin-$1/2$ Heisenberg chain with nearest neighbor interaction. This well-known model has been studied extensively and can be treated exactly via the Bethe-ansatz \cite{bethe:31, baxter:82}. The spin chain including an applied magnetic field was solved with the help of analytical and numerical methods \cite{kluemper:98, takahashi:09}. It will therefore be a good model to test the spin VCA since we can compare our results with the exact solutions by Kl\"umper \cite{kluemper:98}. 

We use the Hamiltonian of the isotropic Heisenberg chain with $N$ sites and periodic boundary conditions 
\begin{eqnarray}
	\o{H} \;=\; \sum_{i} \, h \, \o{S}^{z}_{i} \,+\, \sum_{i} \, J \, \left( \o{S}^{z}_{i} \o{S}^{z}_{i+1} \,+\, \frac{1}{2} \, (\o{S}^{+}_{i} \o{S}^{-}_{i+1} \,+\, \o{S}^{-}_{i} \o{S}^{+}_{i+1}) \right) \; .
	\label{eq:36}
\end{eqnarray}
To apply our cluster approximation,
we tile the chain into clusters with sizes between two and six spins and introduce spatially uniform cluster interactions $\o{J}_{c}^{z}$ and $\o{J}_{c}^{t}$, which act as our variational parameters. In principle we could also attach additional auxiliary sites as seen in figure \ref{fig:1}, thus enlarging
the variational space. We will not use this freedom here as we could not observe a significant improvement
of the results for the spin chain. For each of these cluster systems we need to evaluate the SVCA free energy \eqref{eq:28} 
as described in \ref{app:SVCA}. To this end we use full diagonalization to find the eigenvectors and -values of the cluster Hamiltonian and then compute \eqref{eq:28} 
\begin{equation}
	\beta F_{SVCA}(\o{J}_{c}) \,=\, \beta F_{c} \,+\, K^{z} \,+\, K^{t} \;\;\; ,
	\label{eq:37}
\end{equation}
with $K^{t}:=\Tr \ln \left(\frac{\o{\Gamma}_{c}^{t} \,-\, \o{J}^{t}}{\o{\Gamma}_{c}^{t} \,-\, \o{J}_{c}^{t}}\right)$
and $K^{z}:=\Tr \ln \left(\frac{\o{\Gamma}_{c}^{z} \,-\, \o{J}^{z}}{\o{\Gamma}_{c}^{z} \,-\, \o{J}_{c}^{z}}\right)$. 
Explicit expressions for these quantities are given in \ref{app:SVCA}, equations \eqref{eq:A10} and \eqref{eq:A14}. 

\begin{figure}[tb]
\vspace*{-3mm}
\begin{center}
\includegraphics[width=1\linewidth]{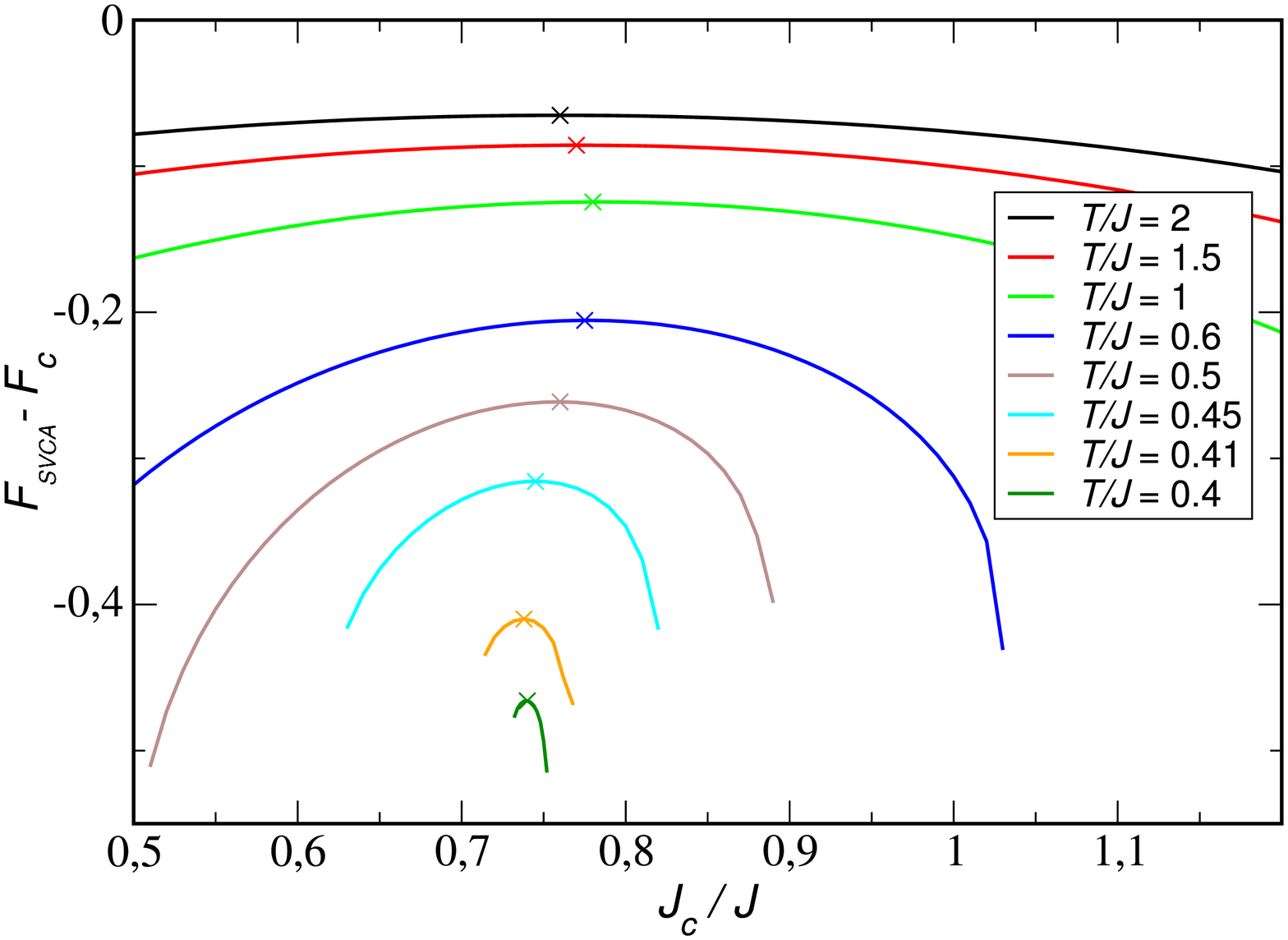}
  \end{center}
\caption{The difference $F_{SVCA}-F_{c}$ between the SVCA and the cluster system free energy per spin according to \eqref{eq:28} for an antiferromagnetic spin chain using two-site clusters. It is plotted as a function of the intra-cluster exchange interaction $J_{c}/J$ for $h/J=1$ and several temperatures. The maxima of the curves are indicated explicitly.}
  \label{fig:2}
\end{figure}

As an example we show in Fig.\ \ref{fig:2} the results of an evaluation of the SVCA free energy for a two-site cluster and a magnetic field $h/J=1$. The cluster interaction is chosen to be isotropic, so the only variational parameter is $J_{c}^{z}=J_{c}^{t}\equiv J_{c}$. To obtain a better picture of the relevant features the difference $F_{SVCA}-F_{c}=T (K^{t} \,+\, K^{z})$ is plotted.
Some characteristic properties can be discussed with the help of this plot. In VCA approaches one generally 
searches for extremal points of thermodynamical functionals respectively functions with respect to the variational 
quantities. 
It can be easily seen in Fig.\ \ref{fig:2} that extremal points exist for some temperatures, at least down to 
$T=0.4J$. 
With decreasing temperature we observe that the algorithm does not give meaningful
results for certain regimes of $J_{c}$  in accordance with the arguments in the previous section. 
In such cases no SVCA free energy is plotted in Fig.\ \ref{fig:2}.
Note that the regions where the algorithm
breaks down become more and more extended as the temperature decreases,
until we finally do not find any reasonable solutions to the SVCA any more. When this is happening for certain parameters the approximation as a whole fails. In the present example this takes place around $T=0.4J$. 
For those regions where the algorithm works, we obtain smooth curves with well developed extremal points,
which are actually maxima. This is  a direct consequence of the derivation of Eq.\ \eqref{eq:28}. 
We note that in the results shown in Fig.\ \ref{fig:2} we do not observe a significant dependence on the k-mesh, but encounter
such a problem when we reach the parameter regime where the present algorithm does not converge any more.

Once the free energy $F_{SVCA}(J_{c})$ has been calculated, we need to determine the location $\hat{J}_{c}$ of the maxima. Within the present implementation of the algorithm this can be done with an accuracy of order of $10^{-3}$, which also provides an
estimate of the numerical error of the computation. Within the SVCA the extremal point $F_{SVCA}(\hat{J}_{c})$ provides an approximation to the physical free energy of the system. We can use it to derive other thermodynamical quantities, for example the magnetization as the derivative $\frac{\partial F}{\partial h}$, which is performed numerically via the central difference. By scanning the parameter space we can then determine these thermodynamical quantities as functions of the temperature $T$
and magnetic field. 

\begin{figure}[tb]
\centering
\includegraphics[width=1\linewidth]{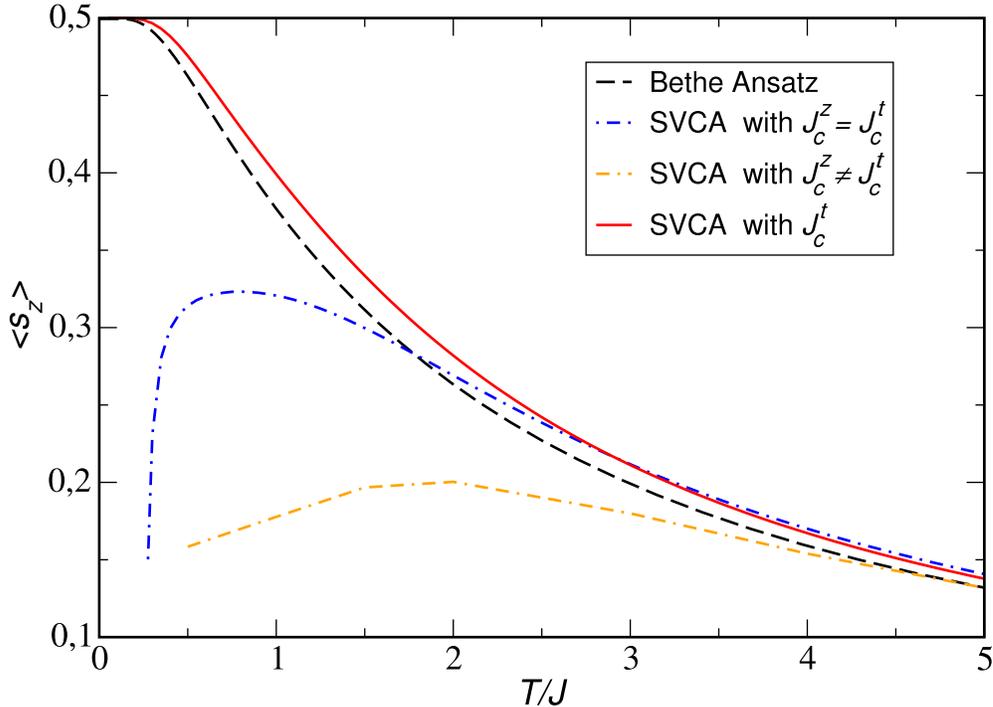}
\caption{The magnetization per spin as function of $T/J$ for the antiferromagnetic spin chain with magnetic field $h=3J$. The exact Bethe Ansatz solution \cite{kluemper:98} is compared with SVCA results for a two-site cluster system and different choices of variational parameters. The low-temperature behavior of the second and third plot is omitted for convenience.}
\label{fig:3}
\end{figure}

In Fig.\ \ref{fig:3} the magnetization per spin for a magnetic field $h=3J>h_{sat}=2J$ obtained from a two-site SVCA is plotted versus temperature.
As can be seen from the exact solution derived with the Bethe ansatz the magnetization saturates for small $T$ \cite{kluemper:98}. As will be discussed later we do not expect a breakdown of the SVCA approximation for this large magnetic field. Fig.\ \ref{fig:3} compares the exact solution with our results for several choices of variational parameters. We tested the isotropic case $J_{c}^{t}=J_{c}^{z}$ we already discussed above as well as the case where $J_{c}^{t}\neq J_{c}^{z}$ are two independent parameters. The latter means that we have to find an extremal point in a two-dimensional parameter space, which naturally is numerically more challenging. For the third choice presented in Fig.\ \ref{fig:3} we set $J_{c}^{z}$ constant as $J$ and only vary the transversal interaction $J_{c}^{t}$. 

Obviously, this latter selection of variational parameter yields the best result when compared to the exact solution. The isotropic and even more so the two-dimensional anisotropic variation lead to obviously wrong magnetization curves, showing a suppression of $\langle S_z\rangle$ for lower temperatures. 
The reason for this behavior is that when we use $J_{c}^{z}$ as independent variational parameter
the extremal points of $F_{SVCA}$ appear at values up to $J_{c}^{z}\approx 3.5J$ for low temperatures, while 
 $J_{c}^{t}$  stays between $0.6J$ and $0.8J$. 
As a consequence the approximation is dominated by artificially enhanced antiferromagnetic correlations counteracting the external magnetic field, which can be clearly seen in Fig.\ \ref{fig:3}. 
In the isotropic case, i.e.\ if we vary the two parameters together,  the transversal part holds the longitudinal part at bay. 
Here values range between $J_{c}\approx0.8J$ for $T=5J$ and $J_{c}\approx1.5J$ for low $T$. If we only vary $J_{c}^{t}$ the extremal points of $F_{SVCA}$ are found at values that always lie below $0.9J$, which obviously leads to a better approximation of the spin chain thermodynamics. 

It may at first seem odd that restricting the variational degrees of freedom results in a better representation of the physics.
It can however be understood if one remembers that at least for one-dimensional spin models  the spin operators can be 
mapped onto fermionic operators by the Jordan-Wigner transformation \cite{jordan:28}. 
Under this mapping the transversal terms $\o{S}^{+}\o{S}^{-}$ become the kinetic energy of the new Hamiltonian while the longitudinal terms $\o{S}^{z}\o{S}^{z}$ lead to density-density interactions of the fermions. For the fermionic version of the VCA
on the other hand it turns out that variational parameters connected to the kinetic energy lead to sensible approximations, while  the interaction part should be kept fixed.

Another aspect of the behavior of the variational parameter $J_{c}^{z}$ is in our opinion related to the magnetic field $h$ applied in the $z$-direction. As we stated above we cannot use $h$ as a variational parameter,
i.e.\ it has to remain fixed. 
In the fermionic or bosonic version of the VCA local fields can be used  to enforce thermodynamical consistency between cluster and real system, for example with respect to
the occupation number in electronic systems \cite{aichhorn:06, senechal:08}. 
As in the present formulation of the SVCA we do not
have such direct control over the magnetization in $z$-direction through a variation of local magnetic fields, the SVCA seems
to maximize the effect of the fixed external field by strongly increasing the longitudinal interaction $J_{c}^{z}$. This behavior is generally encountered in our SVCA computations for the spin chain. 
In the following we will thus present results solely with $J_{c}^{z}=J$ fixed, using $J_{c}^{t}$ as the variational parameter.

\begin{figure}[tp]
\centering
\includegraphics[width=1 \linewidth]{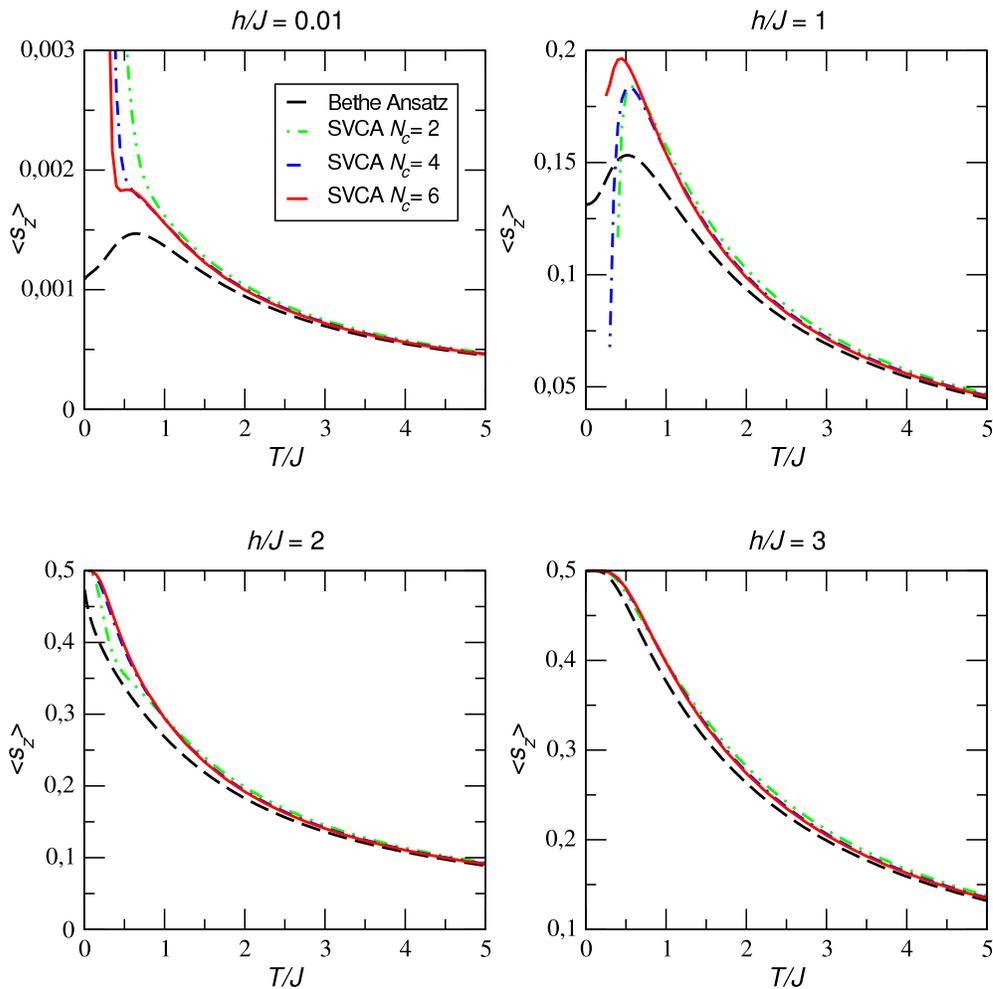}
\caption{The magnetization per spin as a function of the temperature derived with the SVCA for four different magnetic fields. In each graph the results for three cluster choices are plotted together with the exact Bethe ansatz solution.}
\label{fig:4}
\end{figure}

Figure \ref{fig:4} shows the magnetization curves as function of temperature for four different magnetic fields, 
including the critical $h_{sat}=2J$. For larger fields the magnetization per spin is known to smoothly saturate to $1/2$ \cite{griffiths:64}. This region is represented in Fig.\ \ref{fig:4} by a plot for $h=3J$. When $h<2J$ the magnetization curves go through a maximum to settle for a finite value less than $1/2$ as $T\to0$. Two graphs show results for magnetic fields below the critical field. Each plot in
Fig.\ \ref{fig:4} contains the exact Bethe ansatz solution and SVCA results for three different cluster sizes, namely two-, four- and six-site clusters. We choose only even numbers of spins because a dangling spin for odd number of sites prohibits singlet formation in the individual cluster which leads to an expectedly unreliable approximation of the antiferromagnetic spin chain at least for small $h$. 

As can be seen in Fig. \ref{fig:4}, the SVCA approximation works best for the largest magnetic field, $h=3J$. This holds true in general for any $h>2J$, where the magnetization saturates. In this case we also find that the approximation remains stable
down to $T=0$. The saturation value of $1/2$ is found numerically with good precision, which is remarkable without some variational local field as a control parameter. This also supports our choice to only vary $J_{c}^{t}$. 
The dependence on cluster size appears to be very mild, the curves for the four- and six-spin systems already nearly coincide. Thus, 
at least in the case of larger magnetic fields reliable approximations can be obtained for small to moderate cluster sizes.

The results for the critical value $h=2J$ seem to behave similarly at first glance, in particular the approach to the value
$1/2$ for $T=0$. However, the specific form of the exact solution
with its non-analyticity as $T\to0$ is not captured properly. 
Moreover, from the kink 
of the magnetization curve for the two-site cluster at low $T$ one can infer the appearance of additional irregular behavior.
This is a general feature in all our SVCA calculations close to the critical field, which is more or less pronounced depending on
the actual quantity under consideration. For example, it is extremely prominent in the specific heat, see below.
Finally, for $h<2J$ our approximation starts to break down at some finite temperature, as can be seen in Fig.\ \ref{fig:4}.
It thus seems that the SVCA indeed realizes that at $h_c$ a special situation arises for the model. 

Therefore, at $h/J=0.01$ and $h/J=1$ zero temperature can not be reached anymore as the SVCA starts to develop
complex poles in the propagators for the whole parameter space and the results become meaningless. We however note that with increasing cluster
size the stability region also increases and solutions are found for lower $T$. Indeed, especially the curves for the four- and six-spin cluster do not deviate much from one another until at a certain point the solution for the smaller system breaks away while the larger one continues to provide reasonable results down to lower $T$, until it starts to
become unstable, too. This behavior shows that the SVCA indeed is an approximation that at least  improves systematically with
cluster size. Note that such a behavior can in principle be expected from the formulation, but is nevertheless by no means
trivial. 

As can be seen in the plots the magnetization usually becomes divergent to either plus or minus infinity when the SVCA  breaks down. 
It is of course tempting to interpret this breakdown as a signal for a phase transition, albeit an artificial one. 
The antiferromagnetic Heisenberg spin chain does not have a true phase transition at any temperatures \cite{mermin:66}, but is critical in the sense that it develops algebraic correlations at $T=0$ \cite{fazekas:99}.
Adding a not too large magnetic field does not change this situation, but only results in a finite magnetization less than $1/2$ \cite{fazekas:99,kluemper:98}. Since mean-field like theories such as the SVCA tend to produce phase transitions if the correlations in the clusters become too strong, we suggest that the SVCA here tries to form an ordered state to accommodate the slow decay of correlations in the cluster. As the analytical structure of the quantities entering the SVCA should be different in such a situation, we cannot expect that our present implementation is suitable to handle it properly. In particular the appearance of
a finite sub-lattice magnetization will lead to intrinsic consistency problems as discussed in the previous section. 

In any case, apparently the SVCA for small clusters is not able to properly describe the 
region where the spins form some correlated, non-saturated state at low temperature.  
However, the curves in Fig. \ref{fig:4} for the six-spin cluster at least show  the maximum of the magnetization present in the exact
solution, although the magnetization values 
for both $h=0.01J$ and $h=J$ are systematically too high, overestimating the exact value in the maximum by $\approx25$\%. 
As mentioned earlier such a behavior should actually be expected generically because we do not have the option
to adjust a local field as control parameter to enforce a certain magnetization value. 
Yet it is interesting that the positions of the maxima are close to their exact values: 
For $h=0.01J$ the SVCA predicts  $T_\text{max,SVCA}\approx 0.55J$ to be compared with $T_\text{max,BA}\approx0.65J$,
while for  $h=J$ we obtain $T_\text{max,SVCA}\approx 0.44J$ versus $T_\text{max,BA}\approx0.5$.
Thus the SVCA with our present setup seems to systematically overestimate the magnetization -- this is also true for larger magnetic
fields, albeit not that strongly -- while it underestimates the fluctuation scale related to the position of the maximum. Note that of course both features are related and the directions they are going  consistent.

\begin{figure}[tp]
\centering
\includegraphics[width=1 \linewidth]{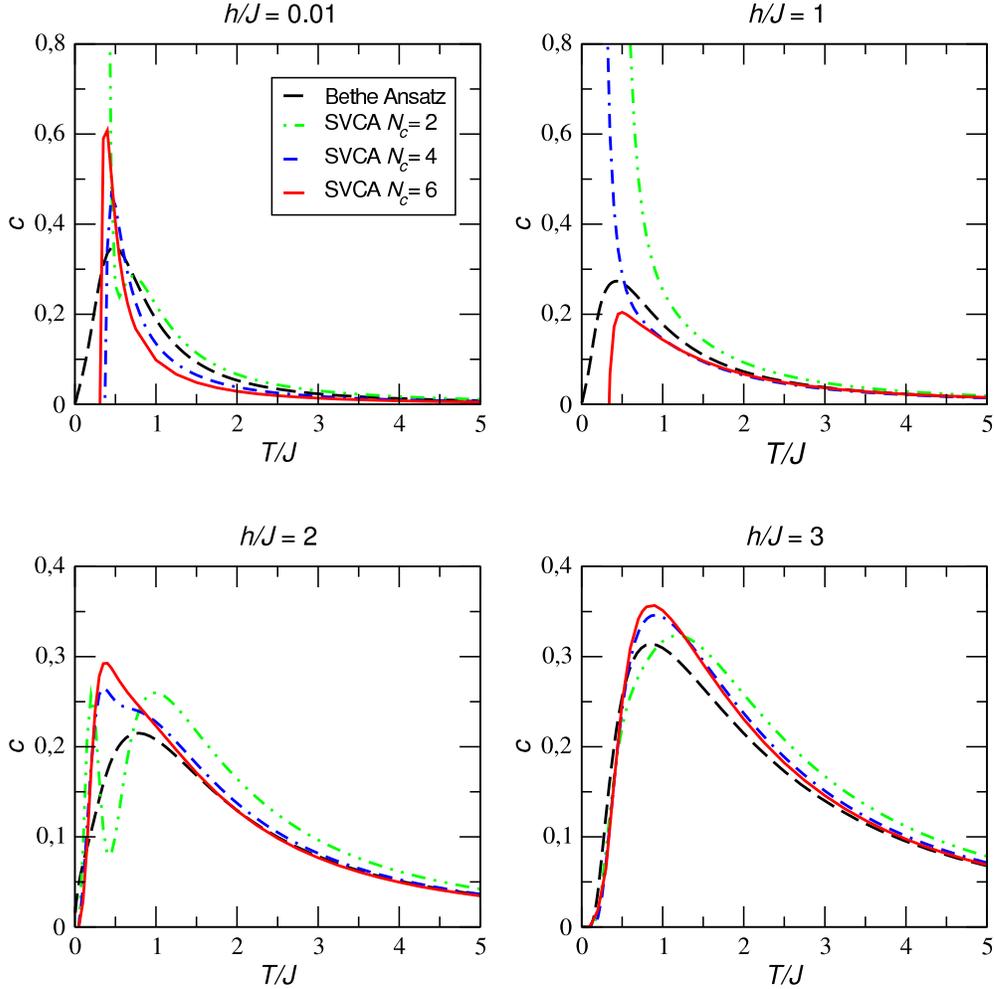}
\caption{\small The heat capacity per spin is shown as a function of the temperature derived with the SVCA for four different magnetic fields. In each graph the results for three cluster choices are plotted together with the exact Bethe ansatz solution.}
\label{fig:5}
\end{figure}

To conclude the discussion of the SVCA results for the antiferromagnetic spin chain we show in Fig.\ \ref{fig:5} our results  for
the specific heat as a further example. The cluster sizes and parameters used  are the same as for the magnetization in Fig.\ \ref{fig:4}.
For the heat capacity $c=-T\frac{\partial^{2} F}{\partial T^{2}}$ holds, i.e.\  one needs to numerically calculate the second
derivative, which is more prone to numerical errors. From the plots in Fig. \ref{fig:5} we directly see that the SVCA results for the heat capacity are less accurate than for the magnetization, even for larger magnetic fields. For the critical value $h=2J$ we find
additional artificial structures at low temperatures. Increasing the cluster size  improves the agreement with the exact Bethe ansatz curve
down to roughly $T\approx1.4J$, but deviations remain significant for lower $T$ and do not seem to improve systematically. 
The situation becomes even worse for $h<2J$ where again divergencies appear and also the overall shape is not reproduced that well
any more. At least for $h=J$ the tendency seems to follow the expectation, viz
that increasing the cluster size yields a systematic improvement of the results, but for $h=0.01J$ even this feature seems
to be lost. 
While one can expect that deficiencies of the approximation as well as numerical errors are more pronounced in quantities which
are obtained as higher derivatives, it is not clear at the moment why the scaling with cluster size of the heat capacity so
significantly deviates from the expected behavior for small magnetic fields.

In the case of a large magnetic field $h=3J$ we again can calculate the specific heat down to $T=0$ and the curves 
approximate the exact solution reasonably well. As expected, the largest cluster provides the best approximation. 
Again, the SVCA results tend to overestimates the heat capacity around the maximum, but predict  its position 
with good accuracy, which is improving with increasing cluster size. 
This confirms our previous observation from the magnetization that for magnetic fields above the critical field the SVCA results 
are more reliable than below $h=2J$.

\section{Summary and Discussion} \label{sec:Discussion}

With the SVCA we present a new cluster approach for Heisenberg spin systems. It is inspired by  the SEFA which
was originally proposed by Potthoff for fermionic systems with local interactions \cite{potthoff:03a}  and which has been subsequently extended to bosonic degrees of freedom \cite{koller:06} and to cluster approximations like Tong's EVCA for more complex models \cite{tong:05}. 
Using a path-integral representation for the partition function of a Heisenberg model we developed a theory that allows
to devise such a variational method for spin models. In the original SEFA, the key object is the single-particle self-energy of
fermions or bosons, and approximations are applied directly to this quantity. For spin models a similar object  is not readily available,
but  by means of spin diagram techniques we could identify the perturbatively defined Larkin self-energy \cite{izyumov:88, pikalev:69} as a suitable quantity for this step of the theory. 
We finally derived a set of variational equations which  yield approximate solutions for the properties of spin systems by introducing certain solvable cluster reference systems and searching for stationary points of the variational free energy $F_{SVCA}$. These approximate values for $F_{SVCA}$ can be used as a starting point to derive other thermodynamical quantities for the spin model under consideration.

There are however certain specific problems which arise and have to be addressed for the SVCA.
One important limitation in the present formulation of the method is that we cannot use local fields as variational parameters.
These variational parameters appear routinely in fermionic or bosonic theories and are actually necessary to control
thermodynamic consistency between reference cluster and physical lattice. 
Moreover, the analytical structure of the SVCA can lead to a failure to find physical solutions in certain situations. This breakdown of the theory can possibly be interpreted as a signal for a phase transition, 
although it need not be one of the physical model. A more conservative point of view is that such a breakdown indicates that
the present choice of cluster or set of  variational parameters is not able to  treat certain properties like an increase of correlation lengths properly. If this is the case, one should be able to see an improvement by choosing larger
clusters or variational parameters that are better suited for the problem at hand. In any case such a breakdown of the SVCA 
can be used to monitor important changes in the physical state of the spin model. 

To test the SVCA we studied the antiferromagnetic Heisenberg $S=1/2$ chain in a magnetic field where we could compare our results with the exact Bethe ansatz solutions. Three reference systems -- full Heisenberg interaction, transversal and longitudinal interaction separately and transversal interaction only as variational parameters -- with different cluster sizes were used for several magnetic fields to apply the SVCA to the model. It turned out that the longitudinal interaction is not suited as variational parameter, probably due to the fact that one cannot use a local field in the cluster Hamiltonians as variational
parameter to control the local magnetization explicitly. 
 
Beyond the critical point $h=2J$ the SVCA results are satisfactory and at least qualitatively resemble the exact solution. For magnetic fields below the critical value we start to encounter a breakdown of the approximation for low temperatures and small
clusters, with a systematic improvement with increasing cluster size, but the results are in general less reliable than for $h>2J$. Accompanied with a breakdown we often find divergencies in the thermodynamical quantities. 
We attribute these to the inability of the used clusters to cope with the algebraic correlations that emerge for low $T$. This interpretation is in accordance with the observation that the results typically become better and more reliable with growing cluster size.

To summarize, we provided a proof of principle that the SVCA can lead to reasonable approximate results for Heisenberg spin models. To this end we chose a simple Heisenberg spin chain, concentrating  on understanding the analytical structure of the
SVCA equations and using a full exact diagonalization to treat the individual clusters. Further investigations with  larger
clusters 
are needed to confirm and extend our observations. However, to achieve this goal one will need more efficient algorithms to
treat open spin clusters and possibly also employ different algorithms to evaluate the quantities entering the SVCA expression
for the free energy .

Of course applying the SVCA to an antiferromagnetic spin chain can only be the first step. To really establish its usefulness
one should also apply it to other models, e.g.\ ladder systems or two-dimensional lattices, including larger spins. Our present
results indicate that the SVCA improves when long-ranged correlations are suppressed. This observation 
makes it particularly interesting to apply the approximation to frustrated spin systems, where short-ranged correlations 
often dominate for low temperatures. For instance, this condition is met in several two-dimensional antiferromagnetic spin lattices \cite{honecker:04}. 
One specific one-dimensional example would be the Heisenberg zig-zag ladder at the Majumdar-Ghosh point, where the
ground state are dimers on the rungs \cite{majumdar:69, mikeska:04}. Work along these lines is in progress.

\ack
We like to acknowledge fruitful discussions with 
M.~Potthoff, N.-H.~Tong and A.~Honecker. 
This work is partially funded (TP) through project P7 of the DFG
research unit FOR 1807.
We acknowledge support by the Open Access Publication Funds of the G\"ottingen University.

\appendix

\section{Evaluation of the Spin VCA Equations}\label{app:SVCA}

To find the stationary points needed for the SVCA one has to evaluate Eq.~\eqref{eq:28}, i.e.
\begin{eqnarray}
	\beta F_{SVCA}(\o{J}_{c}) &=& 
	\beta F_{c} \,+\, \Tr \ln \left(\frac{\o{\Gamma}_{c}^{z} \,-\, \o{J}^{z}}{
	\o{\Gamma}_{c}^{z} \,-\, \o{J}_{c}^{z}}\right) 
	+\, \Tr \ln \left(\frac{\o{\Gamma}_{c}^{t} \,-\, \o{J}^{t}}{
	\o{\Gamma}_{c}^{t} \,-\, \o{J}_{c}^{t}}\right) \endL  \nonumber  \\
	&=& \beta F_{c} \,-\, \Tr \ln \left( \left(\o{\Pi}_{c}^{z}\right)^{-1} \,-\, \o{V}^{z}\right)^{-1} \,+\, \Tr \ln \o{\Pi}_{c}^{z} \nonumber \\
	&& \;\; \,-\, \Tr \ln \left( \left(\o{\Pi}_{c}^{t}\right)^{-1} \,-\, \o{V}^{t}\right)^{-1}  \,+\, \Tr \ln \o{\Pi}_{c}^{t} \;\;\; , 
	\label{eq:A1}
\end{eqnarray}
where we used $\left(\o{\Pi}_{c}^{\xi}\right)^{-1}=\o{\Gamma}_{c}^{\xi} - \o{J}_{c}^{\xi}$ and introduced $\o{V}^{\xi}:= \o{J}^{\xi} - \o{J}_{c}^{\xi}$. All quantities with the subscript $c$ belong to a chosen reference system where our original lattice of $N$ sites is tiled into clusters of $N_{c}$ sites each. 

It is now advisable to Fourier transform the terms in \eqref{eq:A1} with respect to the meta lattice of the clusters \cite{senechal:08}. We end up with a reduced wave vector representation in which the correlation functions $\o{\Pi}_{c}^{\xi}$ are naturally diagonal. The interaction matrix $\o{V}^{\xi}$ on the other hand is not and so will be dependent on a wave vector $\o{k}$. The respective traces in the SVCA free energy \eqref{eq:A1} will thus transform into sums over the cluster site indices and $\o{k}$. In variational cluster approaches the summation over the wave vectors can be approximated by a grid covering the reduced Brillouin zone where the number of terms in the sum is given by the number of clusters $N/N_{c}$. 
In addition to the sums over the lattice indices the traces in \eqref{eq:A1} also include a sum over bosonic Matsubara frequencies $\omega_{n}=2 \pi n T$. To compute the SVCA free energy one now has to provide a suitable and efficient way to carry out the different summations. Several strategies have been introduced for the variational cluster approaches \cite{potthoff:03b, aichhorn:06, aichhorn:06a, koller:06, senechal:08}. 

Before we introduce  the procedure used in the present work we want to note that the contributions to Eq. \eqref{eq:A1} for the longitudinal respectively transversal correlation function are evaluated separately, i.e.~ we will treat the terms
\begin{equation}
	K^{\xi}(\o{J}_{c}) \,:=\, - \Tr \ln \left( \left(\o{\Pi}_{c}^{\xi}\right)^{-1} \,-\, \o{V}^{\xi}\right)^{-1} \,+\, \Tr \ln \o{\Pi}_{c}^{\xi} \,=\, \Tr \ln \left( \o{1} \,-\, \o{V}^{\xi} \o{\Pi}_{c}^{\xi} \right) \;\;\; , 
	\label{eq:A2}
 \end{equation}
individually for $\xi=t$ and $\xi=z$. We will start with the discussion of the transversal part.

In the following we will make use of the $Q$-matrix formalism which was introduced for fermionic \cite{zacher:02, aichhorn:06a} and bosonic systems \cite{koller:06}. There, a Lehmann representation of the corresponding Green function is used as the starting point. We do the same in our case with the cluster spin correlation function. The transversal function $\o{\Pi}^{t}_{c}=\left< \o{S}^{-}_{i}(\tau) \o{S}^{+}_{j}(\tau')  \right>_{c}$ for the cluster can be written using Matsubara frequencies as
\begin{eqnarray}
	\left(\o{\Pi}^{t}_{c}\right)_{i j}(\omega_{l}) &=& \frac{1}{Z} \sum_{n,m} \, \frac{e^{-\beta E_{n}} - e^{-\beta E_{m}}}{\omega_{l} - (E_{n}-E_{m})} \, \left< m | \o{S}^{-}_{i} | n \right> \, \left< n | \o{S}^{+}_{j} | m \right> \nonumber \\
	&=& \frac{1}{Z} \sum_{n,m} \, \frac{e^{-\beta E_{n}} - e^{-\beta E_{m}}}{\omega_{l} - (E_{n}-E_{m})} \, \left< n | \o{S}^{+}_{i} | m \right> \, \left< n | \o{S}^{+}_{j} | m \right>\;\; , 
	\label{eq:A3}
\end{eqnarray} 
where the vectors $\left| n \right>$ are the eigenstates of the cluster system Hamiltonian and the $E_{n}$ the corresponding energies. This object can be analytically continued to establish a function of the complex variable $\omega$. We want to write the correlation function in the $Q$-matrix representation. To this end we define with the help of a multi-index $\alpha = (n,m)$
\begin{eqnarray}
	\o{Q}^{t}_{\alpha i} &=& \sqrt{\frac{\left| e^{-\beta E_{n}} - e^{-\beta E_{m}}\right| }{Z}}
	\left< n | \o{S}^{+}_{i}  | m \right> \newL
	\o{g}_{\alpha \beta}^{t} &=&  \delta_{\alpha \beta} \,  \sgn \left( e^{-\beta E_{n}} - e^{-\beta E_{m}}\right) \newL
	\o{\lambda}_{\alpha \beta}^{t} &=&   \delta_{\alpha \beta} \, (E_{n}-E_{m}) \,=\, \lambda^{t}_{\alpha} \newL
	\o{\Lambda}^{t}(\omega) &=& \frac{\o{g}^{t}}{\omega - \o{\lambda}^{t}} \;\;\; ,
	\label{eq:A4}
\end{eqnarray}
where the $\omega$ in the last line has to be understood as being multiplied by the unity matrix. It is important to note that a certain combination $\alpha$ is taken into account only when the correlation function has a non-vanishing pole at $\lambda_{\alpha}^{t}$. With the definitions in \eqref{eq:A4} we can now write
\begin{equation}
	\o{\Pi}_{c}^{t}(\omega) \,=\, (\o{Q}^{t})^{+} \, \o{\Lambda}^{t}(\omega) \, \o{Q}^{t} \,=\, (\o{Q}^{t})^{+} \, \frac{\o{g}^{t}}{\omega - \o{\lambda}^{t}} \o{Q}^{t} \endL
	\label{eq:A5}
\end{equation}
The other term under the trace in \eqref{eq:A2} can also be rewritten with the above definitions as
\begin{eqnarray}
	 \left( \left(\o{\Pi}_{c}^{t}\right)^{-1} \,-\, \o{V}^{t}\right)^{-1} &=& \o{\Pi}_{c}^{t} \,  \left( \o{1} \,-\, \o{V}^{t} \o{\Pi}_{c}^{t}\right)^{-1} \newL
	 &=&  (\o{Q}^{t})^{+} \, \o{\Lambda}^{t} \, \o{Q}^{t} \,  \left( \o{1} \,-\, \o{V}^{t}  (\o{Q}^{t})^{+} \, \o{\Lambda}^{t} \, \o{Q}^{t} \right)^{-1} \newL
	 &=& (\o{Q}^{t})^{+} \, \o{\Lambda}^{t} \, \left( \o{1} \,-\, \o{Q}^{t} \, \o{V}^{t}  (\o{Q}^{t})^{+} \, \o{\Lambda}^{t} \right)^{-1} \, \o{Q}^{t} \newL
	 &=& (\o{Q}^{t})^{+} \,  \left( (\o{\Lambda}^{t})^{-1} \,-\, \o{Q}^{t} \, \o{V}^{t}  (\o{Q}^{t})^{+} \right)^{-1} \, \o{Q}^{t} \newL
	 &=& (\o{Q}^{t})^{+} \, \o{g}^{t} \, \frac{1}{\omega - \left( \o{\lambda}^{t}  + \o{Q}^{t} \o{V}^{t} (\o{Q}^{t})^{+} \o{g}^{t} \right)} \, \o{Q}^{t} \newL
	 &=& (\o{Q}^{t})^{+} \, \o{g}^{t} \, \o{M}^{-1} \, \frac{1}{\omega - \o{\eta}^{t}(\o{k})} \, \o{M} \, \o{Q}^{t} \;\;\; .
	 \label{eq:A6}
\end{eqnarray}
The step from the second line to the third can be shown by expanding the inverse \cite{aichhorn:06a}. In the last line we introduced the modal matrix $\o{M}$ which diagonalizes $\o{L}^{t}=\left( \o{\lambda}^{t}  + \o{Q}^{t} \o{V}^{t} (\o{Q}^{t})^{+} \o{g}^{t} \right)$. The $\o{\eta}^{t}(\o{k})$ denotes a matrix with the eigenvalues $\eta^{t}_{\alpha}(\o{k})$ on the diagonal. These are wave vector dependent by virtue of the interaction matrix $\o{V}^{t}$. 

Due to its derivation in chapter \ref{sec:SpinVCA} the term \eqref{eq:A6} can be viewed as a matrix containing approximations to the correlation functions of the original system. This view is supported by the general structure we derived in the last line. If we sum over the wave vectors $\o{k}$ the eigenvalues $\eta^{t}_{\alpha}(\o{k})$ represent an approximation to the excitations of the full system. This directly leads to a problem of the present approach. Since $\o{Q}^{t} \o{V}^{t} (\o{Q}^{t})^{+}$ is a hermitian matrix and $\o{g}^{t}$ has varying entries $\pm 1$ the matrix $\o{L}^{t}$ is non-hermitian. This means that one can in principle obtain eigenvalues that lie on the imaginary axis which is not in accordance with their supposed interpretation as physical excitations. The possible imaginary poles will also pose mathematical problems for the evaluation of the SVCA free energy. We will discuss this point below.

The next step is to carry out the traces in \eqref{eq:A2}. 
It is clear that the  Matsubara frequency sum will be the most challenging part. Usually, the correlation function \eqref{eq:A3} decreases relatively fast with $\omega_{l}$. Therefore, for very large $T$ it is sufficient to consider only a finite number of terms. 
For $T \to 0$ on the other hand the sum over the Matsubara frequencies can be carried out efficiently as a numerical integration \cite{senechal:08}. For the present approach however it is mandatory to devise an algorithm working at  arbitrary temperatures, for reasons discussed at the end of section \ref{sec:SpinVCA2.3}.
The analytical technique we will apply was originally introduced for fermions \cite{potthoff:03b} and bosons \cite{koller:06}. In the evaluation of the SVCA we can proceed along the lines of the latter because the correlation function of the spin operators has a structure similar to the bosonic Green function \cite{negele}. 

We start with the term in \eqref{eq:A2} that only incorporates the matrix of cluster correlation functions and is thus independent of the wave vectors $\o{k}$
\begin{eqnarray}
	\Tr \ln \o{\Pi}_{c}^{t} \;=\; \frac {N}{N_{c}} \, \sum_{i, \omega_{n}} \, \ln \pi^{t}_{i}(\omega_{n}) \;\;\; ,
	\label{eq:A7}
\end{eqnarray}
where the $\pi^{t}_{i}(\omega_{n})$ are the eigenvalues of $\o{\Pi}_{c}^{t}$. In evaluating these objects a subtlety arises: each contains a certain number of excitations in such a way that every individual  $\lambda^{t}_{\alpha}$ defined in Eq.~\eqref{eq:A4}
appears only once. This is not trivial to see and cannot be discussed within this paper. It is however important to ensure a proper normalization of the trace.

Next we have to evaluate the individual frequency sums over the terms $\ln \pi^{t}_{i}(\omega_{n})$. A standard way to do this is
by means of Poisson's summation formula. 
An alternative approach suited for the terms appearing in \eqref{eq:A7} is covered in detail by Koller and Dupuis \cite{koller:06}. The result depends solely on the poles $\lambda^{t}_{\alpha}$ and zeros $\zeta^{t}_{\alpha}$ of the functions $\pi^{t}_{i}(\omega)$
\begin{eqnarray}
	\Tr \ln \o{\Pi}_{c}^{t} \;=\; \frac {N}{N_{c}} \, \left( -\sum_{\alpha} \ln \left| 1 - e^{-\beta \, \lambda^{t}_{\alpha}}  \right| \,+\, \sum_{\alpha} \ln \left| 1 - e^{-\beta \, \zeta^{t}_{\alpha}} \right| \right) \; .
	\label{eq:A8}
\end{eqnarray}
Note that the sum over the index $i$ has already been taken into account. For the derivation it is important that the correlation functions are analytical at $\omega=0$. This is  the case for the transversal correlation function 
$\o{\Pi}^{t}$, except for a $SU(2)$-symmetric system. We will handle this point in detail when the longitudinal correlation function is discussed. 

Using a similar computation one finds for the $\o{k}$-dependent part of (\ref{eq:A2})
\begin{eqnarray}
	&& \Tr \ln \left( \left(\o{\Pi}_{c}^{t}\right)^{-1} - \o{V}^{t}\right)^{-1}  \;=\;  \nonumber \\ 
	&& \;\;\;\;\; \;=\; \left( -\sum_{\alpha, \o{k}} \ln \left| 1 - e^{-\beta \, \eta^{t}_{\alpha}(\o{k})}  \right| \,+\, \sum_{\alpha, \o{k}} \ln \left| 1 - e^{-\beta \, \nu^{t}_{\alpha}(\o{k})} \right| \right) \endL
	\label{eq:A9}
\end{eqnarray}

The poles $\eta^{t}_{\alpha}(\o{k})$ were found in \eqref{eq:A6} as the eigenvalues of the matrix $\o{L}^{t}$. The zeros $\nu^{t}_{\alpha}(\o{k})$ on the other hand are determined by the poles of the spin self energy $\o{\Gamma}^{t}$ defined in equation \eqref{eq:13}. In the central approximation of this approach that led to \eqref{eq:28} one effectively uses $\o{\Gamma}^{t}_{c}$ as the self energy for both the cluster as well as the lattice correlation function. So the collection of zeros $\lbrace \zeta^{t}_{\alpha} \rbrace$ is equal to $\lbrace \nu^{t}_{\alpha}(\o{k}) \rbrace$ which means that \eqref{eq:A8} and \eqref{eq:A9} combined leads to
\begin{eqnarray}
	K^{t} \,=\, \sum_{\alpha, \o{k}} \ln \left| 1 - e^{-\beta \, \eta^{t}_{\alpha}(\o{k})}  \right| \,-\, \frac{N}{N_{c}} \, \sum_{\alpha} \ln \left| 1 - e^{-\beta \, \lambda^{t}_{\alpha}}  \right| \endL
	\label{eq:A10}
\end{eqnarray}

To evaluate the longitudinal terms in \eqref{eq:A1} we first express the cluster correlation functions again in their Lehmann representation as
\begin{eqnarray}
	\o{\Pi}^{z}_{i j}(\omega_{l}) &=& \frac{1}{Z} \sum_{E_{n} \neq E_{m}}  \, \frac{e^{-\beta E_{n}} - e^{-\beta E_{m}}}{\omega_{l} - (E_{n}-E_{m})} \, \left< n | \o{S}^{z}_{i} | m \right> \, \left< n | \o{S}^{z}_{j} | m \right> \nonumber \\
	&& \, + \, \delta_{l , 0} \; \frac{\beta}{Z} \, \sum_{E_{m} = E_{n}} \, e^{-\beta E_{n}} \left< n | \o{S}^{z}_{i} | m \right> \, \left< n | \o{S}^{z}_{j} | m \right>
	\;\;\; ,
	\label{eq:A11}
\end{eqnarray}
where the self-adjoint character of $\o{S}^{z}$ has been taken care of and the special contribution to the zeroth component been separated. 

One comment is in order before we proceed with the evaluation. In $K^{z}$ the inverse and the logarithm of $\o{\Pi}^{z}$ appear. It is easy to see that in case total $S^{z}$ is conserved this matrix has zero as an eigenvalue for all Matsubara frequencies except $\omega_{0}$. Thus, $(\o{\Pi}^{z})^{-1}$ and $\ln \o{\Pi}^{z}$ individually have to be in principle understood as containing a small regularizing parameter $\epsilon$ to take care of these zero eigenvalues. It can however be shown that the individual divergencies which develop for $\epsilon \to 0$ exactly cancel if we take into account all terms in $K^{z}$ respectively \eqref{eq:A1}. Therefore the SVCA free energy as a whole is well-defined.
Nevertheless, care has to be taken when one tries to evaluate the different contributions individually.

Let us now proceed with the evaluation of the longitudinal parts. 
The first line of \eqref{eq:A11} resembles terms that also appear in the transversal correlation function \eqref{eq:A3}, and in principle one could perform a similar computation. Unfortunately  the second line proves to be problematic. If we extend \eqref{eq:A11} to be a function on the complex plane it has a discontinuous point at the origin. 
This non-analyticity of $\o{\Pi}^{z}_{i j}(\omega)$ would render it impossible to evaluate the Matsubara frequency sum as in the transversal case. The deformation of the contour leads to a proper result only if the function is analytical at the origin \cite{koller:06}. 

So we introduce a matrix $\hat{\o{\Pi}}_{c}^{z}$ which is defined by \eqref{eq:A11} but with the last line omitted. The corresponding functions are analytical at the origin. Since $\hat{\o{\Pi}}_{c}^{z}$ and $\o{\Pi}_{c}^{z}$ only differ for $\omega_{0}$ we can rewrite
the expressions as
\begin{eqnarray}
	\Tr \ln \o{\Pi}_{c}^{z} \;=\; \Tr \ln \hat{\o{\Pi}}_{c}^{z} \,+\, \Tr \ln \o{\Pi}_{c}^{z} \, |_{\omega_{0}} \,-\, \Tr \ln \hat{\o{\Pi}}_{c}^{z} \, |_{\omega_{0}} \;\;\; ,
	\label{eq:A12}
\end{eqnarray}
and
\begin{eqnarray}
	\Tr \ln \left( \left(\o{\Pi}_{c}^{z}\right)^{-1} \,-\, \o{V}^{z}\right)^{-1} &=& \Tr \ln \left( \left(\hat{\o{\Pi}}_{c}^{z}\right)^{-1} \,-\, \o{V}^{z}\right)^{-1} \nonumber \\
	&& + \; \left. \Tr \ln \left( \left(\o{\Pi}_{c}^{z}\right)^{-1} \,-\, \o{V}^{z}\right)^{-1}\right|_{\omega_{0}} \nonumber \\
	&& - \; \left. \Tr \ln \left( \left(\hat{\o{\Pi}}_{c}^{z}\right)^{-1} \,-\, \o{V}^{z}\right)^{-1}\right|_{\omega_{0}} \; .
	\label{eq:A13}
\end{eqnarray}
The first terms on the right hand side of \eqref{eq:A12} and \eqref{eq:A13} can now be evaluated as in the transversal case, with $\hat{\o{\Pi}}_{c}^{z}$, the $\o{Q}^{z}$-matrices and corresponding quantities like $\o{L}^{z}$ defined in analogy to their transversal
partners. The poles $\lambda_{\alpha}^{z}$ are obtained  from the matrix \eqref{eq:A11} and the $\eta_{\alpha}^{z}(\o{k})$ are determined by the eigenvalues of $\o{L}^{z}$. Both of them are used in the formula corresponding to \eqref{eq:A10}. 

The last two terms from \eqref{eq:A12} and \eqref{eq:A13} which only give a contribution for $\omega_{0}$ can be combined and computed directly. Putting everything together we arrive at
\begin{eqnarray}
	K^{z} &=& \sum_{\alpha, \o{k}} \ln \left| 1 - e^{-\beta \, \eta^{z}_{\alpha}(\o{k})}  \right| \,-\, \frac{N}{N_{c}} \, \sum_{\alpha} \ln \left| 1 - e^{-\beta \, \lambda^{z}_{\alpha}}  \right| \nonumber \\
	&& + \; \sum_{\o{k}} \ln \det \left( \o{1} \,-\, \o{V}^{z} \o{\Pi}_{c}^{z}(\omega_{0}) \right) \nonumber \\
	&& - \; \sum_{\o{k}} \ln \det \left( \o{1} \,-\, \o{V}^{z} \hat{\o{\Pi}}_{c}^{z}(\omega_{0}) \right) \endL
	\label{eq:A14} 
\end{eqnarray}
For the last two terms we used the identity $\Tr \ln \o{M}= \ln \det \o{M}$. The matrices under the determinants cover the cluster site indices. 
One has to add that in case of a $SU(2)$-symmetric system the treatment of the transversal part has to be performed of course identically to the longitudinal part.

Expressions \eqref{eq:A10} and \eqref{eq:A14} together give the SVCA free energy \eqref{eq:A1}. Besides the cluster system free energy $F_{c}$ and excitations $\lambda_{\alpha}^{\xi}$ one has to compute the determinants in \eqref{eq:A14}. One also needs to determine the eigenvalues of the $\o{L}^{\xi}$ which can become numerically challenging. These matrices have a rank of the order of excitations in the system and grow rapidly. Finally, the $\o{k}$-summation appearing in the equations has to be performed and is done over a mesh of different sizes to reduce numerical errors. 

Of course a basic problem is to find the eigenvalues and eigenvectors of the cluster Hamiltonian. We use in this paper full diagonalization, which is limited to smaller clusters. An alternative would be some finite temperature Lanczos method, density-matrix renormalization group or Monte-Carlo approach.

\vspace*{3mm}
\bibliographystyle{unsrt}
\bibliography{references}

\end{document}